\documentclass[manuscript,screen, nonacm]{acmart}
\makeatletter

\AtBeginDocument{%
  \providecommand\BibTeX{{%
    \normalfont B\kern-0.5em{\scshape i\kern-0.25em b}\kern-0.8em\TeX}}}

\copyrightyear{2024}
\acmYear{2024}
\setcopyright {rightsretained}
\acmConference[CHI EA '24]{Extended Abstracts of the CHI Conference on Human Factors in Computing Systems}
{May 11--16, 2024}{Honolulu, HI, USA}

\acmBooktitle{Extended Abstracts of the CHI Conference on Human Factors in Computing Systems (CHI EA '24), May 11--16, 2024, Honolulu, HI, USA}

\acmDOI{10.1145/3613905.3651086}
\acmISBN{979-8-4007-0331-7/24/05}

\begin{document}

\title[Holistic Interface Design for Automated Vehicles]{Exploring Holistic HMI Design for Automated Vehicles: Insights from a Participatory Workshop to Bridge In-Vehicle and External Communication}

\author{Haoyu Dong}
\email{h.dong@tue.nl}
\affiliation{%
  \institution{Eindhoven University of Technology}
  \city{Eindhoven}
  \country{Netherlands}
}
\orcid{0000-0002-7164-7608}

\author{Tram Thi Minh Tran}
\email{tram.tran@sydney.edu.au}
\affiliation{Design Lab, %
  \institution{The University of Sydney}
  \streetaddress{}
  \city{Sydney}
  \country{Australia}
}
\orcid{0000-0002-4958-2465}

\author{Rutger Verstegen}
\email{r.verstegen@tue.nl}
\affiliation{%
  \institution{Eindhoven University of Technology}
  \city{Eindhoven}
  \country{Netherlands}
}
\orcid{0000-0003-1934-2898}

\author{Silvia Cazacu}
\email{silvia.cazacu-bucica@kuleuven.be}
\affiliation{%
  \institution{KU Leuven}
  \streetaddress{}
  \city{Leuven}
  \country{Belgium}
}
\orcid{0000-0002-7952-0919}

\author{Ruolin Gao}
\email{r.gao@tue.nl}
\affiliation{%
  \institution{Eindhoven University of Technology}
  \city{Eindhoven}
  \country{Netherlands}
}
\orcid{0009-0003-8898-5093}

\author{Marius Hoggenmüller}
\email{marius.hoggenmuller@sydney.edu.au}
\affiliation{Design Lab, %
  \institution{The University of Sydney}
  \streetaddress{}
  \city{Sydney}
  \country{Australia}
}
\orcid{0000-0002-8893-5729}

\author{Debargha Dey}
\email{debargha.dey@cornell.edu}
\affiliation{Information Sciences, %
  \institution{Cornell Tech}
  \streetaddress{2 W Loop Rd, NY 10044}
  \city{New York}
  \country{United States}
}
\orcid{0000-0001-9266-0126}

\author{Mervyn Franssen}
\email{m.franssen@tue.nl}
\affiliation{%
  \institution{Eindhoven University of Technology}
  \city{Eindhoven}
  \country{Netherlands}
}
\orcid{0009-0009-3095-4829}

\author{Markus Sasalovici}
\email{markus.sasalovici@uni-ulm.de}
\affiliation{%
  \institution{Institute of Media Informatics, Ulm University}
  \city{Ulm}
  \country{Germany}
}
\orcid{0000-0001-9883-2398}

\author{Pavlo Bazilinskyy}
\email{p.bazilinskyy@tue.nl}
\affiliation{%
  \institution{Eindhoven University of Technology}
  \city{Eindhoven}
  \country{Netherlands}
}
\orcid{0000-0001-9565-8240}

\author{Marieke Martens}
\email{m.h.martens@tue.nl}
\affiliation{
  \institution{Eindhoven University of Technology }
  \city{Eindhoven}
  \country{Netherlands}
}
\orcid{0000-0002-1661-7019}

\renewcommand{\shortauthors}{Dong, et al.}

\begin{abstract} 
Human–Machine Interfaces (HMIs) for automated vehicles (AVs) are typically divided into two categories: internal HMIs for interactions within the vehicle, and external HMIs for communication with other road users. In this work, we examine the prospects of bridging these two seemingly distinct domains. Through a participatory workshop with automotive user interface researchers and practitioners, we facilitated a critical exploration of holistic HMI design by having workshop participants collaboratively develop interaction scenarios involving AVs, in-vehicle users, and external road users. The discussion offers insights into the escalation of interface elements as an HMI design strategy, the direct interactions between different users, and an expanded understanding of holistic HMI design. This work reflects a collaborative effort to understand the practical aspects of this holistic design approach, offering new perspectives and encouraging further investigation into this underexplored aspect of automotive user interfaces.

\end{abstract}


\maketitle

\section{Introduction} 
The integration and acceptance of automated vehicles (AVs) into our transportation systems hinges, amongst other things, upon their ability to communicate effectively. This communication is crucial not only for the occupants of the vehicle, such as drivers and passengers, but also for external road users including pedestrians, cyclists, and drivers of manual vehicles~\cite{Bengler2020HMIs, Hollander2021, dey2020taming, tabone2021vulnerable, colley2022investigating, avsar2021efficient, papakostopoulos2021effect}. In this context, extensive research has been conducted regarding the design of human–machine interfaces (HMIs) for AVs, adopting a reductionist approach~\cite{greenpaperUX2007} that focuses either exclusively on internal interfaces (iHMIs) or external interfaces (eHMIs). 

Bridging this segregation, \citet{Bengler2020HMIs} previously proposed an HMI framework for automated driving. This framework categorises HMIs based on their orientation towards internal and external communication, aligning with the standards outlined in ISO/TR 21959~\cite{iso2019road}. 
Central to this framework lies the emphasis on synchronisation and consistency across different types of HMIs, advocating for a holistic HMI design approach to communication in AVs. While this theoretical work has called for further research on the coordination of internal and external communication, the limited research on this approach raises questions: Is it due to a perceived lack of relevant use cases, or are there inherent challenges in implementing a holistic HMI? This gap in literature necessitates further investigation into the practical implementation and its potential impacts in real-world scenarios.

To address this gap, we conducted a participatory workshop with twelve researchers and practitioners in the field of automotive user interfaces. Our objective was not to assume the necessity of such integration but to facilitate an open and critical exploration of potential use cases and scenarios involving holistic HMIs. 

The workshop resulted in three distinct scenarios showcasing the potential benefits of employing holistic HMI design. It is important to note, however, that holistic HMIs are not positioned as universal solutions for all contexts. The initial insights from our workshop suggest potential applications and opportunities for enhancing user interactions with AVs through holistic HMIs, and discuss notable challenges in this area.

This late-breaking work breaks new ground in the field of automotive HMI design and research by showcasing the promise of holistic HMIs in certain situations. We hope that this can act as a launching pad for discussions around the strategy of taking holistic HMIs into account from the beginning of the design process. This paves the way towards an actionable investigation of an underexplored area of AV interaction.

\section{Participatory Workshop}

A participatory workshop was held as part of an academic conference \anon[(conference name redacted for anonymity)]{AutomotiveUI conference 2023, in Ingolstadt, Germany~\cite{holisticHMIws2023}}.
Twelve participants attended the workshop, all of whom were researchers or practitioners in Human-Computer Interaction (HCI) and human factors, or technology consultancy. They varied in their experience, ranging from junior researchers/ PhD students, to experienced professors or industry professionals. Their research focus lay within automotive user interfaces (iHMI, eHMI, and/or general automotive human factors), which were represented by coloured badges handed out upon arrival.



The workshop started with an introduction of the objectives, schedule, and expected outcomes. Two invited keynote speakers, specialised in iHMIs and eHMIs, then provided an overview of the state of the art in their respective domains. This was followed by first round of plenary discussion, where participants equally voiced their ideas, concerns, or visions, to form a common understanding of holistic HMI design for AVs.

\begin{figure*}[h]
    \centering
\includegraphics[width=0.9\textwidth]{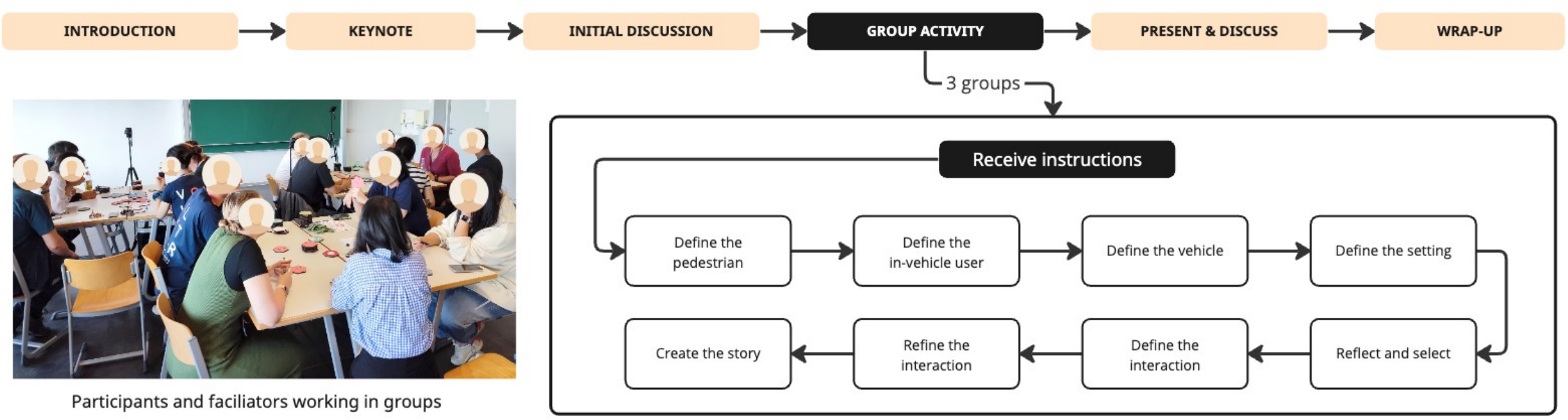}
    \caption{Overview of the workshop.}
    \label{fig:workshop-overview}
    \Description{A diagram illustrating the step-by-step process followed during the workshop, with a photo capturing the scenario of the workshop, depicting participants and facilitators engaged in group activities at three tables.}
\end{figure*} 

Participants were then divided into three groups for the facilitated group activity. Each group consisted of four members with mixed research focus, based on the coloured badges. 
During the group activity, each group was tasked to collaboratively develop one interaction scenario involving multiple traffic participants, thus setting up use cases for holistic HMIs. The workshop concluded with each group presenting their scenarios, followed by a final plenary discussion that reflected on the holistic HMI design approach. The workshop overview is shown in \autoref{fig:workshop-overview}.

\subsection{Group Activity}

To assist participants in creating scenarios in which perspectives of multiple users are considered, we utilised the participatory workshop technique with a set of toolkits including detailed instructions and physical tokens~\cite{Sanders2014ProbesTA}. These instructions broke down the scenario into four key components: \textit{Users}, \textit{Vehicle}, \textit{Environmental Setting}, and \textit{Interaction}. For defining each of these components, we provided four guiding questions. 

\paragraph{Users} 
Two types of users were considered: the in-vehicle user and the pedestrian. The questions for defining each user were inspired by the Empathy Map~\cite{empathymap}, a commonly used tool in design thinking:
\begin{itemize}
    \item Who are you? (e.g., age, gender, job) \item What are you doing? 
    \item What are you perceiving/hearing/seeing/smelling? \item What is your state of mind?
\end{itemize}

\paragraph{Vehicle}
The questions that define the vehicle were designed to allow the participants to freely explore, identify, and specify its properties and/or characteristics by taking the perspective of the non-human traffic participants~\cite{Tomitsch2021}:
\begin{itemize}
    \item What type of vehicle are you? (e.g., passenger car, bus, truck) \item How is your external appearance?
    \item Explain how you can support and communicate with your internal users? 
    \item Explain how you can support and communicate with external road users?
\end{itemize}

\paragraph{Environmental Setting}
The questions aimed at describing the environment were focused on defining the spatial and temporal settings. We also included two main aspects (type of road and weather) to define traffic scenarios based on~\cite{Fratini2023}:
\begin{itemize}
    \item What is the day of the year or season?
    \item What is the time of the day?
    \item What is the location and type of road?
    \item How is the weather at the moment?
\end{itemize}

\paragraph{Interaction}
The questions aimed at defining the interactions encompass four key aspects:
\begin{itemize}
    \item{How would the vehicle and the internal user interact, highlighting the vehicle's advanced features?}
    \item{How would the vehicle and the external user interact, highlighting the vehicle's advanced features?}
    \item{How could the vehicle enable an interaction between internal and external users?}
    \item{What could be a direct interaction between internal and external users?}
\end{itemize}

Each group defined all four components, with the order: Users, Vehicle, Environmental Setting, and Interaction. For each component, participants took turns and each participant answered one question by writing down only keywords on the token and briefly presenting their answers. All the tokens were laid out on the table, facilitating an easier overview and rearrangement. 

Then, each group was required to review all the tokens and their connections to resolve conflicts (e.g., `winter' and `heatwave' as weather components of the proposed scenario, which cannot coexist), thereby enabling the creation of a consistent narrative of the scenario. 
Throughout the group activity, a researcher was present as the facilitator in each group to provide guidance and clarification on the instructions as needed, and the group collaboratively came to a consensus regarding the final narrative of the scenario. A meta facilitator assisted all three groups during the collaborative activity, ensuring that they are in sync with each other and that they follow the protocol.

\subsection{Data Collection and Analysis}
Photographs were taken of the three scenarios developed by the three groups. Additionally, with the participants' verbal consent, all group activities and discussions were captured via video and audio recording. Following the workshop, the facilitators of each group summarised the discussions and the created scenarios. They accomplished this by reviewing and annotating the audio recordings, a method influenced by the concept of `direct analysis' in qualitative research~\cite{POINT2023, parameswaran2020toresearch}. To ensure the reliability of our data, a second reviewer--the facilitator from a different group--was assigned to verify the annotations. Subsequently, the scenarios developed were subjected to collaborative analysis and coding by the authors, leading to the extraction of key insights.

\section{Results}
This section details the scenarios and highlights key points of the group discussions. We created sketches to visualise the created scenarios, focusing on depicting the user and the environment (see \autoref{fig:scenarios}). The list of keywords included in creating each scenario is in Appendix \ref{appendixA}. 

\begin{figure*}[h!]
\centering
\includegraphics[width =\textwidth]{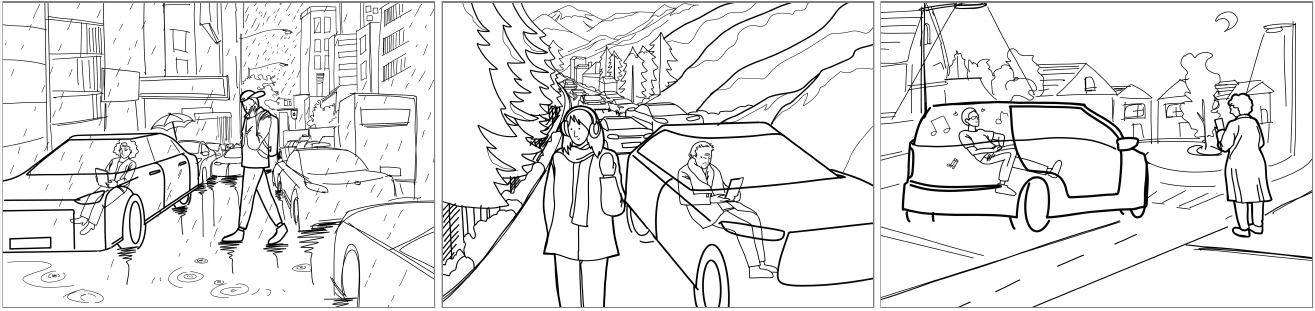}
\caption{Sketches depicting the scenarios: (Left) `Rainy Traffic Jam' scenario, (Middle) `Snowy Mountain Road' scenario, (Right) `Summer Night Roundabout' scenario.}
\label{fig:scenarios}
\Description{Sketches depicting the scenarios draw by hand: (Left) `Rainy Traffic Jam' scenario, (Middle) `Snowy Mountain Road' scenario, (Right) `Summer Night Roundabout' scenario.}
\end{figure*}

\subsection{Scenario One: Rainy Traffic Jam}

\subsubsection{Scenario Description} 

The scenario unfolded in a busy city where a heavy rain caused a traffic jam. Inside the AV, the in-vehicle user remained relaxed, enjoying a YouTube video with the volume turned up. Meanwhile, a stressed pedestrian navigated through the rain, while being engaged in a phone call (as illustrated in \autoref{fig:scenarios} Left).

The AV displayed an eHMI icon with letters, signalling its still-standing traffic jam status. As the pedestrian approached, the AV subtly highlighted the pedestrian's location with lights and non-intrusive audio cues to the in-vehicle user. Similarly, the pedestrian perceived various LED lights and icons through the AV's external display.

There was no direct interaction between the in-vehicle user and the pedestrian, as both were preoccupied with their respective activities.

\subsubsection{Discussion Highlights}
Recognising that both users were distracted from the actual situation engaging in something else, group one contemplated the possibility of intensifying the AV communication to attract both attention, for example, increasing the volume of audio cues, or having the LED light blinking. The intensified HMI may successfully draw both users' attention, fostering a direct interaction between the two, at most a shared glance acknowledging each other's presence. 
Later on, the group considered that the still-standing traffic jam situation might eliminate the necessity of direct interaction and a direct interaction between the internal and external users should be a fallback option in case the AV can not handle a situation.
In this scenario, with no breakdown in the AV's functionality, the absence of direct interaction was deemed acceptable, with both individuals continuing their activities undisturbed.

\subsection{Scenario Two: Snowy Mountain Road}
\subsubsection{Scenario Description} 

In this winter scenario, a mountain road was busy and treacherous due to seasonal traffic and slippery conditions, posing potential hazards. An AV carried an older in-vehicle user, who was absorbed in internet browsing. The AV was equipped with a driver monitoring system, constantly assessing the state of the in-vehicle user. A woman walked along the road and passed by the AV, while listening to a podcast and being mindful of her safety in such challenging conditions. Both users were less alert to their surrounding environment (as illustrated in \autoref{fig:scenarios} Middle).

In the event of danger, the AV employed a transformer-inspired mechanism to alert the in-vehicle user, adapting the warning methods according to the severity of the situation and the driver's current state. This system escalated its alerts from subtle visual signals to auditory warnings and seat vibrations. For external communications, the pedestrian initially received alerts on her smartwatch, functioning as a virtual assistant. If the initial warning was ignored, the system would temporarily interrupt her podcast, ensuring the pedestrian becomes fully attentive to her surroundings.

\subsubsection{Discussion Highlights}
In the interaction defining phase, Group Two considered that both the in-vehicle user and the pedestrian were in a potentially hazardous situation. This realisation influenced the group's decision to focus on guaranteeing safety, by employing uniformity in the escalation of both eHMI and iHMI to ensure both parties are aware of the situation and receive relevant messages. 
This led to a discussion about the potential of simultaneously mirroring information to both parties facing the same danger. However, concerns were also raised about the difference in implementation for iHMI and eHMI (e.g., monitoring systems). Finally, this group discussed the possibility of the vehicle reflecting the internal user's state and emotions, serving as a direct communication channel between the in-vehicle user and the pedestrian. 

\subsection{Scenario Three: Summer Night Roundabout}

\subsubsection{Scenario Description}

This scenario happened on a summer evening with a clear sky. An old woman strolled on the street while watching YouTube videos on her phone. She remained alert to the sounds around her yet not looking around. She approached a roundabout where an autonomous shuttle, reserved for individual use, approached. The shuttle carried a young man, who was eagerly anticipating a date. He was immersed in the music playing from the shuttle's speakers (as illustrated in \autoref{fig:scenarios} Right).

As the shuttle detected the woman,  it subtly adjusted the music volume and activated its virtual avatar to gently notify the in-vehicle user. Simultaneously, the shuttle changed its exterior colour to yellow, in an attempt to alert the woman. However, the woman, absorbed in the Youtube video, remained oblivious. Then, the shuttle extended its outreach beyond its external interface by sending a message to the woman's phone. It also lowered its window, allowing the young man to speak directly to her. 
In adverse weather conditions, this interaction could alternatively occur virtually, with the window remaining closed and the conversation broadcast externally through a speaker inspired by the Tesla Model 3 Boombox\footnote{\url{https://www.tesla.com/ownersmanual/model3/en_us/GUID-79A49D40-A028-435B-A7F6-8E48846AB9E9.html}}.

\subsubsection{Discussion Highlights}

Group Three employed more gentle communication means (e.g., lowering the music, changing the exterior colour) and more noticeable measures (e.g., a talking avatar, text message). They also carefully considered the state and current activities of both users to suggest suitable communication methods in this scenario. The discussion focused on the circumstances surrounding both users—the tranquil evening during which the encounter took place. This context would allow for a more personal form of communication. The positive and non-aggressive states of both individuals also influenced the decision for this interaction. As a result, the final interaction entailed the vehicle opening its window, enabling direct communication between the two users. This scenario illustrates a harmonious interaction facilitated by technology, demonstrating the potential for direct communication between internal and external users in shared spaces.

\section{Discussion} 
In this section, we discuss key similarities and differences among the three scenarios, reflecting on the holistic HMI design approach utilised in developing the scenarios. 

\subsection{HMI Escalation as a Shared Design Strategy}

Regarding the environmental setting, two out of three scenarios involved traffic jams, which is an atypical scenario given that the majority of existing literature on AV communication focuses on fast-paced, high-risk situations~\cite{wilbrink2017interact, tran2021review, Cauffman2022ResearchDriving}. In these slow-moving traffic situations, the kinematic cues of the AV (referred to by~\citet{Bengler2020HMIs} as dynamic HMIs) become harder to observe, potentially necessitating the use of more explicit types of HMIs. Additionally, two out of the three scenarios involved special weather conditions that posed potential hazards, including a snowy mountain road that was narrow and slippery, and a rainy situation that affected the AV sensor performance.

In all scenarios, both the in-vehicle user and the pedestrian were occupied with their own activities, predominantly consuming media. These scenarios mirror real-world situations of distracted pedestrians who use their phones while walking~\cite{hollander2019smombies}. For in-vehicle users, the rise of automated driving systems (ADS) increasingly allows them to engage in non-driving related tasks. 

The influence of environmental settings and the distracted state of the involved users prompted all groups to arrive at a similar strategy of escalation of selective user interface elements in response to non-action of users.  Here, escalation refers to the process of progressively increasing or intensifying the level of interaction between the AV and the human users. 
This escalation is designed to ensure effective communication and response, especially in critical or complex scenarios. Therefore, it was deemed most relevant in scenarios which carried the highest potential risks (e.g., the `Snowy Mountain Road' in our case). 

In implementing this strategy, all the groups considered various modalities and technologies, in some cases, leveraged devices that are sources of engagement or distraction. For example, in the `Summer Night Roundabout' scenario, the shuttle sends a message to the woman's phone and lowers the music inside the shuttle. Moreover, we observed the potential for all related technologies within a traffic scenario to be interconnected, facilitating easier dissemination and optimisation of information delivery. For instance, pedestrians could receive notifications on their own devices. 
With the increase in connected devices and the development of novel systems connecting vehicles (i.e., Vehicle-to-Everything or V2X)~\cite{harding2014vehicle, lu2014connected, gunther2015potential}, the integration of different types of HMIs for a holistic AV communication approach is clearly feasible.

Despite HMI escalation being a shared design strategy that applies to both internal and external communication, the implementation for iHMIs and eHMIs could be different. For example, considerations such as the availability and privacy consent related to monitoring systems, both internally and externally, were discussed by Group Two and are evident in existing literature~\cite{dey2020taming}. 
In particular, an iHMI can often follow a more standardised approach since it deals primarily with the functionality of the car itself, which tends to be more universal. In contrast, eHMIs interact with a broader environment and various road users. This interaction requires a deeper understanding of local customs and non-verbal communication cues~\cite{weber2019crossing, ranasinghe2020culture}. 
As a result, while HMI escalation can be applied to both iHMIs and eHMIs, it may not necessarily have to occur simultaneously or in the same manner for both. This strategy represents a relatively unexplored area that offers significant opportunities for advancing HMI research.

\subsection{Interaction Between In-Vehicle User and Pedestrian}
Interaction between internal and external users, either directly or mediated by the AV, is scarcely considered in the design of HMIs in the context of automated driving. The design and research on iHMI often focus on either input or output channels between the in-vehicle user and the AV, through interfaces with various modalities~\cite{Detjen2021HowReview}. 
Meanwhile, eHMI research typically concerns fully autonomous vehicles without any occupants inside (SAE Level 5~\cite{sae2021taxonomy}). However, insightful findings do exist, such as potential conflicts arising from opposing cues given by drivers or passengers and the eHMI~\cite{colley2022gesticulation}. 

In the workshop, we found varying degrees of interaction between the in-vehicle user and the pedestrian being discussed across the three scenarios. 
The range of interaction varies from no interaction needed (or at most a shared glance) in `Rainy Traffic Jam', to mediated interaction (vehicle expressing the driver's emotion) in `Snowy Mountain Road', to a direct interaction (conversation between the shuttle passengers and pedestrians) in `Summer Night'. 
Regarding the AV expressing the driver's emotions, this aspect echoes with an eHMI dimension referred to as Vehicle Occupant State by~\citet{dey2020taming}, which captures whether the eHMI enables the vehicle to communicate the state of its occupants to external users (e.g., `angst'). 
Besides, the direct interaction was not due to a failure of AV communication, as in a study by~\citet{brown2023halting} where the passenger had to apologise for the AV behaviour, saying \textit{`Sorry, it's a self-driving car.'} Instead, the direct interaction was facilitated by the shuttle lowering its window and adding another layer of interaction, which might aid safety and efficiency. 

\subsection{Towards an Expanded Understanding of Holistic HMI Design}

\citet{Bengler2020HMIs} refers to a holistic HMI communication approach as \textit{`considering all HMI types when researching the interaction strategies of AVs with its passenger or surrounding human road users'}. Findings from our workshop contribute to a more expanded understanding of holistic HMI design. The holistic perspective could imply either a singular design for all users, or an integration of various HMI designs into a cohesive set of interactions.
First, it may involve a shared design strategy that could be applicable for both iHMIs and eHMIs, facilitating information mirroring and unified interaction strategy (e.g., HMI escalation) for consistent communication among all involved parties. 
Second, it also encourages a design process that considers both internal and external users within the same interaction scenario, fostering the integration of multiple designs. 

Contrary to the traditional separation of iHMI and eHMI under a Design-as-Engineering approach~\cite{Wright2006} or a reductionist approach~\cite{greenpaperUX2007}, this holistic perspective aligns with HCI's evolving focus from usability to experience-focused design~\cite{Wright2006, reimaginingperspective2011, Karapanos2013}. 
This shift acknowledges that experience design not only involves the designed system, but also considers user's internal states and the context in which interactions occur~\cite{Hassenzahl2008, Hassenzahl2010}. 

In the context of AVs, this holistic approach, which serves as a bridge between iHMI and eHMI, underscores the importance of integrating both the AV and internal and external users into the same setup. This integration is vital for creating cohesive user experiences and emphasising how AVs mediate and alter human users' activities and perceptions in daily life.
We posit that the design space of HMI for AV shows the potential of expanding to the design of an `interspace' (proposed by~\citet{winograd1997}) inhabited by multiple people and AVs, in a traffic environment with complex interactions. This view also aligns with research focus on scalability in HMI design for AV~\cite{dey2020taming, tran2023scoping}. 

Furthermore, the holistic approach acknowledges the intricate interconnections among various factors that shape user experience, without sacrificing complexity for easy measurements of the impact of individual HMI elements~\cite{UXmanifesto2007, Karapanos2013, Ackoff1973reductionist, greenpaperUX2007}. 
This perspective underscores the importance of a coherent design language capable of accommodating the dynamic roles individuals assume in diverse traffic environments. For instance, users may seamlessly transition between roles as pedestrians, passengers, or drivers in their daily life, experiencing either iHMI or eHMI at different time points. This necessitates the implementation of adaptable interfaces.

\section{Limitations and Future Work}

Despite effort to mix participants with diverse backgrounds in the group activities, noticeable similarities emerged in the scenarios developed by all three groups. This observation raises the possibility of a convergence in thought process or a general agreement in the research community when approaching HMI design for AVs. This shared bias could indicate either a widespread tendency in the domain of automotive HMI, or could be attributed to the design of the group activities. Hence, while not the primary focus of this paper, it is crucial to contemplate the methodology's potential impact on the final outcomes. Subsequent work will provide a more comprehensive examination of the methodology, offering detailed insights into the design process of the group activity and the participatory workshop toolkit. 
Additionally, given the exploratory nature of this workshop, the scenarios were constrained to include only one pedestrian, one in-vehicle user, and one vehicle. Future efforts should extend to incorporate multiple users, offering a more comprehensive perspective that mirrors the intricate and diverse nature of real-world traffic situations.

By showcasing three scenarios developed during the workshop, the early insights highlight the potential benefits of holistic HMI design, indicating its positive impact on shaping interactions with AVs and elevating user experiences in specific scenarios. 
The findings underscore the viability of such an approach, highlighting the need for a comprehensive exploration of scenarios and use cases where holistic HMI approaches could offer significant value in automotive HMI design. 
Work is underway to elaborate on the scenarios and identify opportunities and challenges within the design space of holistic HMI. This involves multiple brainstorming sessions, and future co-creating workshops with a wider range of specialists to identify such scenarios. We posit that such an exhaustive exploration of applicable scenarios also promises a deeper understanding of the holistic HMI design approach. 
Furthermore, we plan to conduct interviews with experts in the field to gain insights into the multifaceted definition and refine the framework of holistic HMI design approach. By shedding light on potential limitations and challenges, we contribute to future implementations and unlock its full potential in shaping the future of human–vehicle interaction.

\section{Conclusion}
This paper presents three scenarios created at a workshop implementing holistic HMI design approach to bridge internal and external communication in AVs. 
The initial insights suggest the potential of such an approach in enriching interactions with AV and enhancing user experience in specific contexts. Concerns are also raised, highlighting that this is a complex topic, encompassing both promises and challenges—thereby necessitating further exploration.  
Our findings contribute to an expanded understanding of holistic HMI design approach, emphasising a design process early on focusing on the intricate dynamics of the `interspace' where interactions unfold among multiple participants, including in-vehicle users, pedestrians, and AVs. By sharing these preliminary findings within the HCI community, our goal is to catalyse meaningful discussions on the applications of holistic HMI design approach. This serves as a foundation for actionable plans in future work within the relatively under-explored area of human–vehicle interaction.

\begin{acks}
We would like to thank all the workshop participants for their enthusiastic and original work, and all the contributors to make this workshop happen. This research is supported by China Scholarship Council award (No. 202007720018), the Australian Research Council (ARC) Discovery Project DP200102604 (Trust and Safety in Autonomous Mobility Systems: A Human-centred Approach), the Dutch Research Council NWO-NWA Grant (No. NWA.1292.19.298), project ODECO (which has received funding from the European Union’s Horizon 2020 research and innovation programme under the Marie Skłodowska-Curie grant agreement No. 955569), China Scholarship Council award (No. 202006030027).

\end{acks}

\bibliographystyle{Example/ACM-Reference-Format}
\bibliography{reference}


\begin{thebibliography}{38}


\ifx \showCODEN    \undefined \def \showCODEN     #1{\unskip}     \fi
\ifx \showDOI      \undefined \def \showDOI       #1{#1}\fi
\ifx \showISBNx    \undefined \def \showISBNx     #1{\unskip}     \fi
\ifx \showISBNxiii \undefined \def \showISBNxiii  #1{\unskip}     \fi
\ifx \showISSN     \undefined \def \showISSN      #1{\unskip}     \fi
\ifx \showLCCN     \undefined \def \showLCCN      #1{\unskip}     \fi
\ifx \shownote     \undefined \def \shownote      #1{#1}          \fi
\ifx \showarticletitle \undefined \def \showarticletitle #1{#1}   \fi
\ifx \showURL      \undefined \def \showURL       {\relax}        \fi
\providecommand\bibfield[2]{#2}
\providecommand\bibinfo[2]{#2}
\providecommand\natexlab[1]{#1}
\providecommand\showeprint[2][]{arXiv:#2}

\bibitem[iso(2019)]%
        {iso2019road}
 \bibinfo{year}{2019}\natexlab{}.
\newblock \bibinfo{booktitle}{\emph{Road Vehicles: Human Performance and State in the Context of Automated Driving: Part 2--Considerations in Designing Experiments to Investigate Transition Processes}}.
\newblock \bibinfo{type}{ISO/PRF TR} 21959-2. \bibinfo{institution}{International Organization for Standardization}, \bibinfo{address}{Geneva, Switzerland}.
\newblock


\bibitem[Ackoff(1973)]%
        {Ackoff1973reductionist}
\bibfield{author}{\bibinfo{person}{Russell~L. Ackoff}.} \bibinfo{year}{1973}\natexlab{}.
\newblock \showarticletitle{Science in the Systems Age: Beyond IE, OR, and MS}.
\newblock \bibinfo{journal}{\emph{Oper. Res.}}  \bibinfo{volume}{21} (\bibinfo{year}{1973}), \bibinfo{pages}{661--671}.
\newblock


\bibitem[Avsar et~al\mbox{.}(2021)]%
        {avsar2021efficient}
\bibfield{author}{\bibinfo{person}{H{\"u}seyin Avsar}, \bibinfo{person}{Fabian Utesch}, \bibinfo{person}{Marc Wilbrink}, \bibinfo{person}{Michael Oehl}, {and} \bibinfo{person}{Caroline Schie{\ss}l}.} \bibinfo{year}{2021}\natexlab{}.
\newblock \showarticletitle{Efficient communication of automated vehicles and manually driven vehicles through an external Human-Machine Interface (eHMI): Evaluation at T-junctions}. In \bibinfo{booktitle}{\emph{HCI International 2021-Posters: 23rd HCI International Conference, HCII 2021, Virtual Event, July 24--29, 2021, Proceedings, Part III 23}}. Springer, \bibinfo{pages}{224--232}.
\newblock


\bibitem[Bannon(2011)]%
        {reimaginingperspective2011}
\bibfield{author}{\bibinfo{person}{Liam Bannon}.} \bibinfo{year}{2011}\natexlab{}.
\newblock \showarticletitle{Reimagining HCI: Toward a More Human-Centered Perspective}.
\newblock  (\bibinfo{year}{2011}).
\newblock
\urldef\tempurl%
\url{https://doi.org/10.1145/1978822.1978833}
\showDOI{\tempurl}


\bibitem[Bengler et~al\mbox{.}(2020)]%
        {Bengler2020HMIs}
\bibfield{author}{\bibinfo{person}{Klaus Bengler}, \bibinfo{person}{Michael Rettenmaier}, \bibinfo{person}{Nicole Fritz}, {and} \bibinfo{person}{Alexander Feierle}.} \bibinfo{year}{2020}\natexlab{}.
\newblock \showarticletitle{From HMI to HMIs: Towards an HMI framework for automated driving}.
\newblock \bibinfo{journal}{\emph{Information (Switzerland)}}  \bibinfo{volume}{11} (\bibinfo{date}{2} \bibinfo{year}{2020}).
\newblock
Issue 2.
\showISSN{20782489}
\urldef\tempurl%
\url{https://doi.org/10.3390/info11020061}
\showDOI{\tempurl}


\bibitem[Blythe et~al\mbox{.}(2007)]%
        {greenpaperUX2007}
\bibfield{author}{\bibinfo{person}{Mark Blythe}, \bibinfo{person}{Marc Hassenzahl}, \bibinfo{person}{Effie Law}, {and} \bibinfo{person}{Arnold Vermeeren}.} \bibinfo{year}{2007}\natexlab{}.
\newblock \bibinfo{title}{An analysis framework for user experience (UX)studies: a green paper}.
\newblock
\newblock


\bibitem[Brown et~al\mbox{.}(2023)]%
        {brown2023halting}
\bibfield{author}{\bibinfo{person}{Barry Brown}, \bibinfo{person}{Mathias Broth}, {and} \bibinfo{person}{Erik Vinkhuyzen}.} \bibinfo{year}{2023}\natexlab{}.
\newblock \showarticletitle{The Halting Problem: Video Analysis of Self-Driving Cars in Traffic}. In \bibinfo{booktitle}{\emph{Proceedings of the 2023 CHI Conference on Human Factors in Computing Systems}} (Hamburg, Germany) \emph{(\bibinfo{series}{CHI '23})}. \bibinfo{publisher}{Association for Computing Machinery}, \bibinfo{address}{New York, NY, USA}, Article \bibinfo{articleno}{12}, \bibinfo{numpages}{14}~pages.
\newblock
\showISBNx{9781450394215}
\urldef\tempurl%
\url{https://doi.org/10.1145/3544548.3581045}
\showDOI{\tempurl}


\bibitem[Cauffman et~al\mbox{.}(2022)]%
        {Cauffman2022ResearchDriving}
\bibfield{author}{\bibinfo{person}{Stephen~J. Cauffman}, \bibinfo{person}{Mei Lau}, \bibinfo{person}{Yulin Deng}, \bibinfo{person}{Christopher Cunningham}, \bibinfo{person}{David~B. Kaber}, {and} \bibinfo{person}{Jing Feng}.} \bibinfo{year}{2022}\natexlab{}.
\newblock \showarticletitle{{Research and Design Considerations for Presentation of Non-Safety Related Information via In-Vehicle Displays during Automated Driving}}.
\newblock \bibinfo{journal}{\emph{Applied Sciences (Switzerland)}} \bibinfo{volume}{12}, \bibinfo{number}{20} (\bibinfo{date}{10} \bibinfo{year}{2022}).
\newblock
\showISSN{20763417}
\urldef\tempurl%
\url{https://doi.org/10.3390/app122010538}
\showDOI{\tempurl}


\bibitem[Colley et~al\mbox{.}(2022a)]%
        {colley2022investigating}
\bibfield{author}{\bibinfo{person}{Mark Colley}, \bibinfo{person}{Tim Fabian}, {and} \bibinfo{person}{Enrico Rukzio}.} \bibinfo{year}{2022}\natexlab{a}.
\newblock \showarticletitle{Investigating the Effects of External Communication and Automation Behavior on Manual Drivers at Intersections}.
\newblock \bibinfo{journal}{\emph{Proc. ACM Hum.-Comput. Interact}}  \bibinfo{volume}{6} (\bibinfo{year}{2022}), \bibinfo{pages}{1--16}.
\newblock


\bibitem[Colley et~al\mbox{.}(2022b)]%
        {colley2022gesticulation}
\bibfield{author}{\bibinfo{person}{Mark Colley}, \bibinfo{person}{Bastian Wankmüller}, \bibinfo{person}{Tim Mend}, \bibinfo{person}{Thomas Väth}, \bibinfo{person}{Enrico Rukzio}, {and} \bibinfo{person}{Jan Gugenheimer}.} \bibinfo{year}{2022}\natexlab{b}.
\newblock \showarticletitle{User gesticulation inside an automated vehicle with external communication can cause confusion in pedestrians and a lower willingness to cross}.
\newblock \bibinfo{journal}{\emph{Transportation Research Part F: Traffic Psychology and Behaviour}}  \bibinfo{volume}{87} (\bibinfo{year}{2022}), \bibinfo{pages}{120--137}.
\newblock
\showISSN{1369-8478}
\urldef\tempurl%
\url{https://doi.org/10.1016/j.trf.2022.03.011}
\showDOI{\tempurl}


\bibitem[Dam and Teo(2023)]%
        {empathymap}
\bibfield{author}{\bibinfo{person}{R.~F. Dam} {and} \bibinfo{person}{Y.~S. Teo}.} \bibinfo{year}{2023}\natexlab{}.
\newblock \showarticletitle{Empathy Map – Why and How to Use It}.
\newblock \bibinfo{journal}{\emph{Interaction Design Foundation - IxDF}} (\bibinfo{year}{2023}).
\newblock
\urldef\tempurl%
\url{https://www.interaction-design.org/literature/article/empathy-map-why-and-how-to-use-it}
\showURL{%
\tempurl}


\bibitem[Detjen et~al\mbox{.}(2021)]%
        {Detjen2021HowReview}
\bibfield{author}{\bibinfo{person}{Henrik Detjen}, \bibinfo{person}{Sarah Faltaous}, \bibinfo{person}{Bastian Pfleging}, \bibinfo{person}{Stefan Geisler}, {and} \bibinfo{person}{Stefan Schneegass}.} \bibinfo{year}{2021}\natexlab{}.
\newblock \showarticletitle{{How to Increase Automated Vehicles’ Acceptance through In-Vehicle Interaction Design: A Review}}.
\newblock \bibinfo{journal}{\emph{International Journal of Human-Computer Interaction}} \bibinfo{volume}{37}, \bibinfo{number}{4} (\bibinfo{year}{2021}), \bibinfo{pages}{308--330}.
\newblock
\showISSN{15327590}
\urldef\tempurl%
\url{https://doi.org/10.1080/10447318.2020.1860517}
\showDOI{\tempurl}


\bibitem[Dey et~al\mbox{.}(2020)]%
        {dey2020taming}
\bibfield{author}{\bibinfo{person}{Debargha Dey}, \bibinfo{person}{Azra Habibovic}, \bibinfo{person}{Andreas L{\"o}cken}, \bibinfo{person}{Philipp Wintersberger}, \bibinfo{person}{Bastian Pfleging}, \bibinfo{person}{Andreas Riener}, \bibinfo{person}{Marieke Martens}, {and} \bibinfo{person}{Jacques Terken}.} \bibinfo{year}{2020}\natexlab{}.
\newblock \showarticletitle{Taming the eHMI jungle: A classification taxonomy to guide, compare, and assess the design principles of automated vehicles' external human-machine interfaces}.
\newblock \bibinfo{journal}{\emph{Transportation Research Interdisciplinary Perspectives}}  \bibinfo{volume}{7} (\bibinfo{year}{2020}), \bibinfo{pages}{100174}.
\newblock


\bibitem[Dong et~al\mbox{.}(2023)]%
        {holisticHMIws2023}
\bibfield{author}{\bibinfo{person}{Haoyu Dong}, \bibinfo{person}{Tram Thi~Minh Tran}, \bibinfo{person}{Pavlo Bazilinskyy}, \bibinfo{person}{Marius Hoggenm\"{u}ller}, \bibinfo{person}{Debargha Dey}, \bibinfo{person}{Silvia Cazacu}, \bibinfo{person}{Mervyn Franssen}, {and} \bibinfo{person}{Ruolin Gao}.} \bibinfo{year}{2023}\natexlab{}.
\newblock \showarticletitle{Holistic HMI Design for Automated Vehicles: Bridging In-Vehicle and External Communication}. In \bibinfo{booktitle}{\emph{Adjunct Proceedings of the 15th International Conference on Automotive User Interfaces and Interactive Vehicular Applications}} (Ingolstadt, Germany) \emph{(\bibinfo{series}{AutomotiveUI '23 Adjunct})}. \bibinfo{publisher}{Association for Computing Machinery}, \bibinfo{address}{New York, NY, USA}, \bibinfo{pages}{237–240}.
\newblock
\showISBNx{9798400701122}
\urldef\tempurl%
\url{https://doi.org/10.1145/3581961.3609837}
\showDOI{\tempurl}


\bibitem[Fratini et~al\mbox{.}(2023)]%
        {Fratini2023}
\bibfield{author}{\bibinfo{person}{Elena Fratini}, \bibinfo{person}{Ruth Welsh}, {and} \bibinfo{person}{Pete Thomas}.} \bibinfo{year}{2023}\natexlab{}.
\newblock \showarticletitle{Ranking Crossing Scenario Complexity for eHMIs Testing: A Virtual Reality Study}.
\newblock \bibinfo{journal}{\emph{Multimodal Technologies and Interaction}} \bibinfo{volume}{7}, \bibinfo{number}{2} (\bibinfo{year}{2023}).
\newblock
\showISSN{2414-4088}
\urldef\tempurl%
\url{https://doi.org/10.3390/mti7020016}
\showDOI{\tempurl}


\bibitem[Gunther et~al\mbox{.}(2015)]%
        {gunther2015potential}
\bibfield{author}{\bibinfo{person}{Hendrik-jorn Gunther}, \bibinfo{person}{Oliver Trauer}, {and} \bibinfo{person}{Lars Wolf}.} \bibinfo{year}{2015}\natexlab{}.
\newblock \showarticletitle{The potential of collective perception in vehicular ad-hoc networks}. In \bibinfo{booktitle}{\emph{2015 14th International Conference on ITS Telecommunications (ITST)}}. IEEE, \bibinfo{pages}{1--5}.
\newblock


\bibitem[Harding et~al\mbox{.}(2014)]%
        {harding2014vehicle}
\bibfield{author}{\bibinfo{person}{John Harding}, \bibinfo{person}{Gregory Powell}, \bibinfo{person}{Rebecca Yoon}, \bibinfo{person}{Joshua Fikentscher}, \bibinfo{person}{Charlene Doyle}, \bibinfo{person}{Dana Sade}, \bibinfo{person}{Mike Lukuc}, \bibinfo{person}{Jim Simons}, \bibinfo{person}{Jing Wang}, {et~al\mbox{.}}} \bibinfo{year}{2014}\natexlab{}.
\newblock \bibinfo{booktitle}{\emph{Vehicle-to-vehicle communications: readiness of V2V technology for application.}}
\newblock \bibinfo{type}{{T}echnical {R}eport}. \bibinfo{institution}{United States. National Highway Traffic Safety Administration}.
\newblock


\bibitem[Hassenzahl(2008)]%
        {Hassenzahl2008}
\bibfield{author}{\bibinfo{person}{Marc Hassenzahl}.} \bibinfo{year}{2008}\natexlab{}.
\newblock \showarticletitle{User Experience (UX): Towards an experiential perspective on product quality}.
\newblock \bibinfo{journal}{\emph{IHM 2008}}.
\newblock
\showISBNx{9781605582856}


\bibitem[Hassenzahl(2010)]%
        {Hassenzahl2010}
\bibfield{author}{\bibinfo{person}{Marc Hassenzahl}.} \bibinfo{year}{2010}\natexlab{}.
\newblock \showarticletitle{Experience Design: Technology for All the Right Reasons}.
\newblock \bibinfo{journal}{\emph{Synthesis Lectures on Human-Centered Informatics}}  \bibinfo{volume}{3} (\bibinfo{date}{1} \bibinfo{year}{2010}), \bibinfo{pages}{1--95}.
\newblock
Issue 1.
\showISSN{1946-7680}
\urldef\tempurl%
\url{https://doi.org/10.2200/s00261ed1v01y201003hci008}
\showDOI{\tempurl}


\bibitem[Hollander et~al\mbox{.}(2021)]%
        {Hollander2021}
\bibfield{author}{\bibinfo{person}{Kai Hollander}, \bibinfo{person}{Mark Colley}, \bibinfo{person}{Enrico Rukzio}, {and} \bibinfo{person}{Andreas Butz}.} \bibinfo{year}{2021}\natexlab{}.
\newblock \showarticletitle{{A taxonomy of vulnerable road users for hci based on a systematic literature review}}. In \bibinfo{booktitle}{\emph{Conference on Human Factors in Computing Systems - Proceedings}}.
\newblock
\showISBNx{9781450380966}
\urldef\tempurl%
\url{https://doi.org/10.1145/3411764.3445480}
\showDOI{\tempurl}


\bibitem[Holl\"{a}nder et~al\mbox{.}(2020)]%
        {hollander2019smombies}
\bibfield{author}{\bibinfo{person}{Kai Holl\"{a}nder}, \bibinfo{person}{Andy Kr\"{u}ger}, {and} \bibinfo{person}{Andreas Butz}.} \bibinfo{year}{2020}\natexlab{}.
\newblock \showarticletitle{Save the Smombies: App-Assisted Street Crossing}. In \bibinfo{booktitle}{\emph{22nd International Conference on Human-Computer Interaction with Mobile Devices and Services}} (Oldenburg, Germany) \emph{(\bibinfo{series}{MobileHCI '20})}. \bibinfo{publisher}{Association for Computing Machinery}, \bibinfo{address}{New York, NY, USA}, Article \bibinfo{articleno}{22}, \bibinfo{numpages}{11}~pages.
\newblock
\showISBNx{9781450375160}
\urldef\tempurl%
\url{https://doi.org/10.1145/3379503.3403547}
\showDOI{\tempurl}


\bibitem[Karapanos(2013)]%
        {Karapanos2013}
\bibfield{author}{\bibinfo{person}{Evangelos Karapanos}.} \bibinfo{year}{2013}\natexlab{}.
\newblock \bibinfo{booktitle}{\emph{Modeling Users' Experiences with Interactive Systems}}.
\newblock
\showISBNx{978-3-642-30999-1}
\showISSN{1860-949x}
\urldef\tempurl%
\url{https://doi.org/10.1007/978-3-642-31000-3}
\showDOI{\tempurl}


\bibitem[Lu et~al\mbox{.}(2014)]%
        {lu2014connected}
\bibfield{author}{\bibinfo{person}{Ning Lu}, \bibinfo{person}{Nan Cheng}, \bibinfo{person}{Ning Zhang}, \bibinfo{person}{Xuemin Shen}, {and} \bibinfo{person}{Jon~W Mark}.} \bibinfo{year}{2014}\natexlab{}.
\newblock \showarticletitle{Connected vehicles: Solutions and challenges}.
\newblock \bibinfo{journal}{\emph{IEEE internet of things journal}} \bibinfo{volume}{1}, \bibinfo{number}{4} (\bibinfo{year}{2014}), \bibinfo{pages}{289--299}.
\newblock


\bibitem[Papakostopoulos et~al\mbox{.}(2021)]%
        {papakostopoulos2021effect}
\bibfield{author}{\bibinfo{person}{Vassilis Papakostopoulos}, \bibinfo{person}{Dimitris Nathanael}, \bibinfo{person}{Evangelia Portouli}, {and} \bibinfo{person}{Angelos Amditis}.} \bibinfo{year}{2021}\natexlab{}.
\newblock \showarticletitle{Effect of external HMI for automated vehicles (AVs) on drivers’ ability to infer the AV motion intention: A field experiment}.
\newblock \bibinfo{journal}{\emph{Transportation research part F: traffic psychology and behaviour}}  \bibinfo{volume}{82} (\bibinfo{year}{2021}), \bibinfo{pages}{32--42}.
\newblock


\bibitem[Parameswaran et~al\mbox{.}(2020)]%
        {parameswaran2020toresearch}
\bibfield{author}{\bibinfo{person}{Uma~D Parameswaran}, \bibinfo{person}{Jade~L Ozawa-Kirk}, {and} \bibinfo{person}{Gwen Latendresse}.} \bibinfo{year}{2020}\natexlab{}.
\newblock \showarticletitle{To live (code) or to not: A new method for coding in qualitative research}.
\newblock \bibinfo{journal}{\emph{Qualitative Social Work}} \bibinfo{volume}{19}, \bibinfo{number}{4} (\bibinfo{year}{2020}), \bibinfo{pages}{630--644}.
\newblock
\urldef\tempurl%
\url{https://doi.org/10.1177/1473325019840394}
\showDOI{\tempurl}
\showeprint{https://doi.org/10.1177/1473325019840394}


\bibitem[Point and Baruch(2023)]%
        {POINT2023}
\bibfield{author}{\bibinfo{person}{Sébastien Point} {and} \bibinfo{person}{Yehuda Baruch}.} \bibinfo{year}{2023}\natexlab{}.
\newblock \showarticletitle{(Re)thinking transcription strategies: Current challenges and future research directions}.
\newblock \bibinfo{journal}{\emph{Scandinavian Journal of Management}} \bibinfo{volume}{39}, \bibinfo{number}{2} (\bibinfo{year}{2023}), \bibinfo{pages}{101272}.
\newblock
\showISSN{0956-5221}
\urldef\tempurl%
\url{https://doi.org/10.1016/j.scaman.2023.101272}
\showDOI{\tempurl}


\bibitem[Ranasinghe et~al\mbox{.}(2020)]%
        {ranasinghe2020culture}
\bibfield{author}{\bibinfo{person}{Champika Ranasinghe}, \bibinfo{person}{Kai Holl\"{a}nder}, \bibinfo{person}{Rebecca Currano}, \bibinfo{person}{David Sirkin}, \bibinfo{person}{Dylan Moore}, \bibinfo{person}{Stefan Schneegass}, {and} \bibinfo{person}{Wendy Ju}.} \bibinfo{year}{2020}\natexlab{}.
\newblock \showarticletitle{Autonomous Vehicle-Pedestrian Interaction Across Cultures: Towards Designing Better External Human Machine Interfaces (EHMIs)}. In \bibinfo{booktitle}{\emph{Extended Abstracts of the 2020 CHI Conference on Human Factors in Computing Systems}} (Honolulu, HI, USA) \emph{(\bibinfo{series}{Chi Ea '20})}. \bibinfo{publisher}{Association for Computing Machinery}, \bibinfo{address}{New York, NY, USA}, \bibinfo{pages}{1–8}.
\newblock
\showISBNx{9781450368193}
\urldef\tempurl%
\url{https://doi.org/10.1145/3334480.3382957}
\showDOI{\tempurl}


\bibitem[SAE(2021)]%
        {sae2021taxonomy}
\bibfield{author}{\bibinfo{person}{SAE}.} \bibinfo{year}{2021}\natexlab{}.
\newblock \showarticletitle{Taxonomy and definitions for terms related to driving automation systems for on-road motor vehicles}.
\newblock \bibinfo{journal}{\emph{J3016, SAE International}} (\bibinfo{year}{2021}).
\newblock
\urldef\tempurl%
\url{https://doi.org/10.4271/J3016_202104}
\showURL{%
\tempurl}


\bibitem[Sanders and Stappers(2014)]%
        {Sanders2014ProbesTA}
\bibfield{author}{\bibinfo{person}{Elizabeth B.-N. Sanders} {and} \bibinfo{person}{Pieter~Jan Stappers}.} \bibinfo{year}{2014}\natexlab{}.
\newblock \showarticletitle{Probes, toolkits and prototypes: three approaches to making in codesigning}.
\newblock \bibinfo{journal}{\emph{CoDesign}}  \bibinfo{volume}{10} (\bibinfo{year}{2014}), \bibinfo{pages}{14 -- 5}.
\newblock
\urldef\tempurl%
\url{https://api.semanticscholar.org/CorpusID:108955372}
\showURL{%
\tempurl}


\bibitem[Tabone et~al\mbox{.}(2021)]%
        {tabone2021vulnerable}
\bibfield{author}{\bibinfo{person}{Wilbert Tabone}, \bibinfo{person}{Joost De~Winter}, \bibinfo{person}{Claudia Ackermann}, \bibinfo{person}{Jonas B{\"a}rgman}, \bibinfo{person}{Martin Baumann}, \bibinfo{person}{Shuchisnigdha Deb}, \bibinfo{person}{Colleen Emmenegger}, \bibinfo{person}{Azra Habibovic}, \bibinfo{person}{Marjan Hagenzieker}, \bibinfo{person}{Peter~A Hancock}, {et~al\mbox{.}}} \bibinfo{year}{2021}\natexlab{}.
\newblock \showarticletitle{Vulnerable road users and the coming wave of automated vehicles: Expert perspectives}.
\newblock \bibinfo{journal}{\emph{Transportation research interdisciplinary perspectives}}  \bibinfo{volume}{9} (\bibinfo{year}{2021}), \bibinfo{pages}{100293}.
\newblock


\bibitem[Tomitsch et~al\mbox{.}(2021)]%
        {Tomitsch2021}
\bibfield{author}{\bibinfo{person}{Martin Tomitsch}, \bibinfo{person}{Joel Fredericks}, \bibinfo{person}{Dan Vo}, \bibinfo{person}{Jessica Frawley}, {and} \bibinfo{person}{Marcus Foth}.} \bibinfo{year}{2021}\natexlab{}.
\newblock \showarticletitle{Non-human Personas: Including Nature in the Participatory Design of Smart Cities}.
\newblock \bibinfo{journal}{\emph{Interaction Design and Architecture(s)}} (\bibinfo{date}{12} \bibinfo{year}{2021}), \bibinfo{pages}{102--130}.
\newblock
Issue 50.
\showISSN{22832998}
\urldef\tempurl%
\url{https://doi.org/10.55612/s-5002-050-006}
\showDOI{\tempurl}


\bibitem[Tran et~al\mbox{.}(2021)]%
        {tran2021review}
\bibfield{author}{\bibinfo{person}{Tram Thi~Minh Tran}, \bibinfo{person}{Callum Parker}, {and} \bibinfo{person}{Martin Tomitsch}.} \bibinfo{year}{2021}\natexlab{}.
\newblock \showarticletitle{A Review of Virtual Reality Studies on Autonomous Vehicle–Pedestrian Interaction}.
\newblock \bibinfo{journal}{\emph{IEEE Transactions on Human-Machine Systems}} \bibinfo{volume}{51}, \bibinfo{number}{6} (\bibinfo{year}{2021}), \bibinfo{pages}{641--652}.
\newblock
\urldef\tempurl%
\url{https://doi.org/10.1109/thms.2021.3107517}
\showDOI{\tempurl}


\bibitem[Tran et~al\mbox{.}(2023)]%
        {tran2023scoping}
\bibfield{author}{\bibinfo{person}{Tram Thi~Minh Tran}, \bibinfo{person}{Callum Parker}, {and} \bibinfo{person}{Martin Tomitsch}.} \bibinfo{year}{2023}\natexlab{}.
\newblock \showarticletitle{Scoping Out the Scalability Issues of Autonomous Vehicle-Pedestrian Interaction}. In \bibinfo{booktitle}{\emph{Proceedings of the 15th International Conference on Automotive User Interfaces and Interactive Vehicular Applications}} (Ingolstadt, Germany) \emph{(\bibinfo{series}{AutomotiveUI '23})}. \bibinfo{publisher}{Association for Computing Machinery}, \bibinfo{address}{New York, NY, USA}, \bibinfo{pages}{167–177}.
\newblock
\urldef\tempurl%
\url{https://doi.org/10.1145/3580585.3607167}
\showDOI{\tempurl}


\bibitem[Weber et~al\mbox{.}(2019)]%
        {weber2019crossing}
\bibfield{author}{\bibinfo{person}{Florian Weber}, \bibinfo{person}{Ronee Chadowitz}, \bibinfo{person}{Kathrin Schmidt}, \bibinfo{person}{Julia Messerschmidt}, {and} \bibinfo{person}{Tanja Fuest}.} \bibinfo{year}{2019}\natexlab{}.
\newblock \showarticletitle{Crossing the Street Across the Globe: A Study on the Effects of eHMI on Pedestrians in the US, Germany and China}. In \bibinfo{booktitle}{\emph{International Conference on Human-Computer Interaction}}. Springer, \bibinfo{pages}{515--530}.
\newblock


\bibitem[Wilbrink et~al\mbox{.}(2017)]%
        {wilbrink2017interact}
\bibfield{author}{\bibinfo{person}{Marc Wilbrink} {et~al\mbox{.}}} \bibinfo{year}{2017}\natexlab{}.
\newblock \bibinfo{title}{{D1.1 Definition of interACT use cases and scenarios}}.
\newblock
\newblock
\urldef\tempurl%
\url{https://elib.dlr.de/116445}
\showURL{%
\tempurl}


\bibitem[Winograd(1997)]%
        {winograd1997}
\bibfield{author}{\bibinfo{person}{Terry Winograd}.} \bibinfo{year}{1997}\natexlab{}.
\newblock \bibinfo{booktitle}{\emph{The design of interaction}}.
\newblock \bibinfo{publisher}{Copernicus}, \bibinfo{address}{USA}, \bibinfo{pages}{149–161}.
\newblock
\showISBNx{0387949321}


\bibitem[Wright and Blythe(2007)]%
        {UXmanifesto2007}
\bibfield{author}{\bibinfo{person}{Peter Wright} {and} \bibinfo{person}{Mark Blythe}.} \bibinfo{year}{2007}\natexlab{}.
\newblock \showarticletitle{User Experience Research as an Inter-discipline: Towards a UX Manifesto}, \bibfield{editor}{\bibinfo{person}{E.Law}, \bibinfo{person}{A.~Vermeeren}, \bibinfo{person}{M.~Hassenzahl}, {and} \bibinfo{person}{M.~Blythe}} (Eds.).
\newblock \bibinfo{journal}{\emph{Towards a UX Manifesto - Proceedings of a cost294-affiliated workshop on BHCI 2007}}, \bibinfo{pages}{65--70}.
\newblock


\bibitem[Wright et~al\mbox{.}(2006)]%
        {Wright2006}
\bibfield{author}{\bibinfo{person}{Peter Wright}, \bibinfo{person}{Mark Blythe}, {and} \bibinfo{person}{John McCarthy}.} \bibinfo{year}{2006}\natexlab{}.
\newblock \showarticletitle{User experience and the idea of design in HCI}.
\newblock \bibinfo{journal}{\emph{Lecture Notes in Computer Science}}  \bibinfo{volume}{3941 LNCS}, \bibinfo{pages}{1--14}.
\newblock
\showISBNx{3540341455}
\showISSN{16113349}
\urldef\tempurl%
\url{https://doi.org/10.1007/11752707_1}
\showDOI{\tempurl}


\end{thebibliography}

\appendix

\section{Set of Keywords}
\label{appendixA}

\begin{figure*}[h!]
    \centering
    \includegraphics[width =0.7\textwidth]{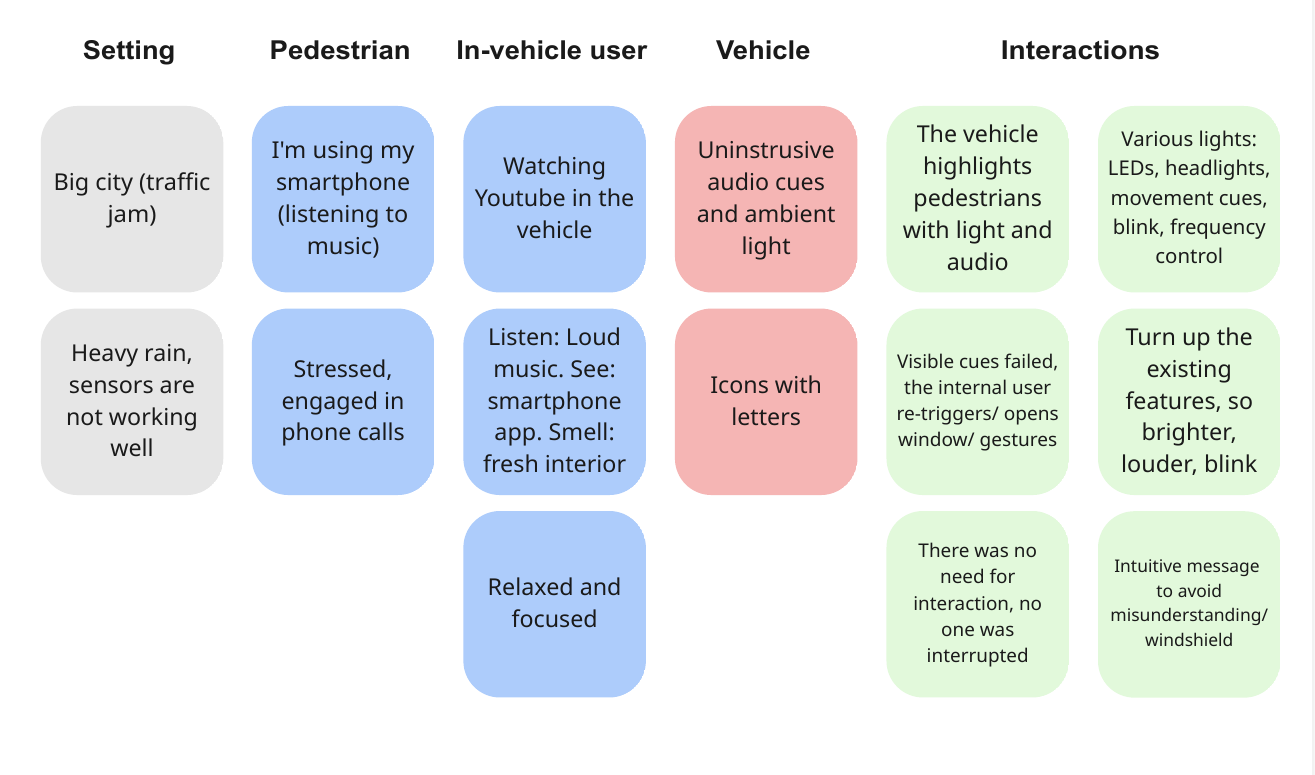}
    \caption{Set of keywords for `Rainy Traffic Jam' scenario (Group One)}
    \label{fig:s1}
    \Description{A figure showing the set of keywords for `Rainy Traffic Jam' scenario (Group One): 
        The first column `Setting' includes two sticky notes: `Big city (traffic jam)' and `Heavy rain, sensors are not working well'. 
    The second column `Pedestrian' includes two sticky notes: `Using my smartphone (listening to music)' and `Feeling stressed, engaged in phone calls'. 
        The third column `In-vehicle user' includes three sticky notes: `Watching YouTube in the vehicle', `Listening to loud music via smartphone app' and `Feeling relaxed and focused in a fresh interior environment'. 
        The fourth column `Vehicle' includes descriptors such as `Unintrusive audio cues and ambient light' and `Icons with letters'. 
       The fifth column `Interactions' includes: `Highlighting pedestrians with light and audio', `Re-triggering internal user actions due to failed visible cues', `Utilizing various lights and movement cues', `Enhancing existing features for clarity', `Providing intuitive messages on the windshield', `No need for interaction, ensuring no interruption to users'.}
\end{figure*}

\begin{figure*}[h!]
    \centering
    \includegraphics[width =0.7\textwidth]{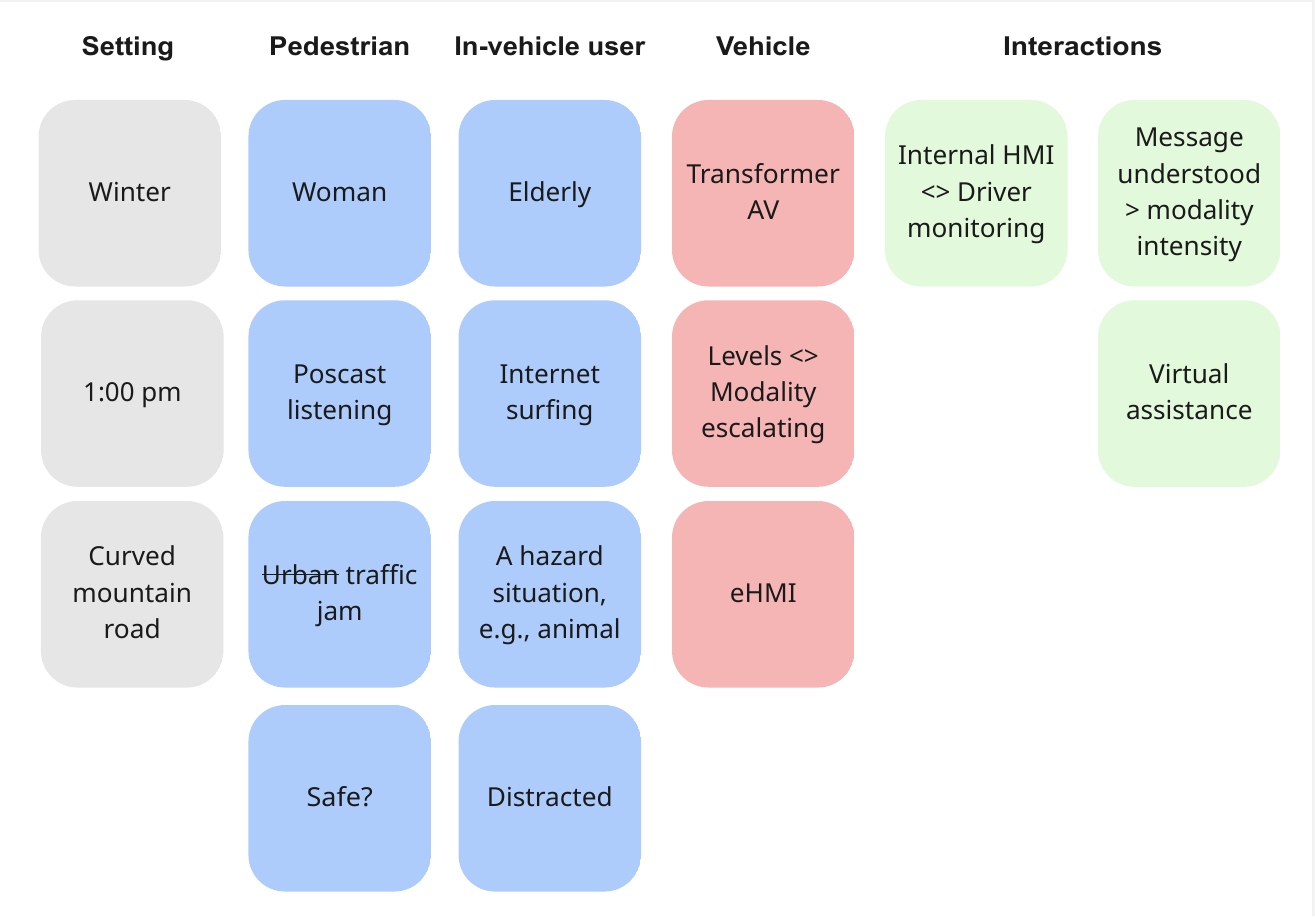}
    \caption{Set of keywords for `Snowy Mountain Road' scenario (Group Two)}
    \label{fig:s2}
    \Description{A figure showing set of keywords for `Snowy Mountain Road' scenario (Group Two):
    The first column `Setting' includes three sticky notes: `Winter', `1:00 pm' and `Curved mountain road'.
    The second column `Pedestrian' includes four sticky notes: `Woman', `Podcast listening', `(Strike-through: Urban) traffic jam', and `Safe?'.
    The third column `In-vehicle user' includes four sticky notes: `Elderly',`Internet surfing',`A hazard situation, e.g. animal', `Distracted'. 
    The fourth column `Vehicle' includes three sticky notes: `Transformer AV', `Levels <> Modality escalating', `eHMI'. 
    The fifth column `Interactions' includes three notes: `Internal HMI <> Driver monitoring',`Message understood > modality intensity', `Virtual assistance'.}
\end{figure*}

\begin{figure*}[h!]
    \centering
    \includegraphics[width =0.7\textwidth]{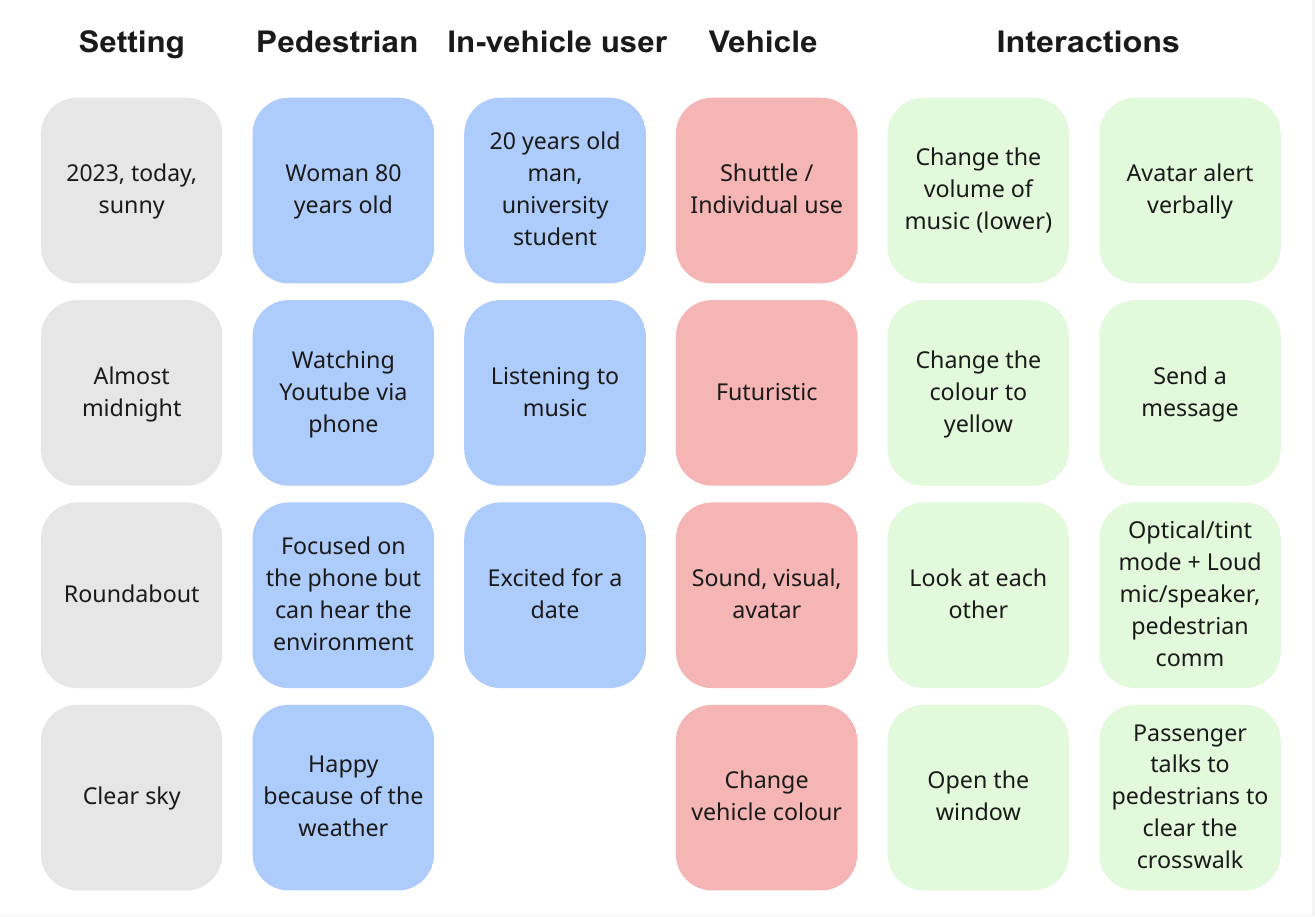}
    \caption{Set of keywords for `Summer Night Roundabout' scenario (Group Three)}
    \label{fig:s3}
    \Description{A figure showing set of keywords for `Summer Night Roundabout' scenario (Group Three):
   The first column `Setting' includes four sticky notes: `2023, today, sunny', `Almost midnight', `Roundabout', `Clear sky'.
        The second column `Pedestrian' includes four sticky notes: `Woman 80 years old',`Watching Youtube via phone',`Focused on the phone but can hear the environment', `Happy because of the weather'.
     The third column `In-vehicle user' includes three sticky notes: `20 years old man, university student',`Listening to music', `Excited for a date'. 
    The fourth column `Vehicle' includes four sticky notes: `Shuttle/Individual use', `Futuristic', `Sound, visual, avatar', `Change vehicle colour'. 
    Lastly, the fifth column `Interactions' includes eight notes: `Change the volume of music (lower)', `Avatar alert verbally', `Change the colour to yellow', `Send a message', `Look at each other', `Optical/tint mode + Loud mic/speaker, pedestrian comm', `Open the window',`Passenger talks to pedestrians to clear the crosswalk'.
    }
\end{figure*}

\end{document}